\documentclass[sigconf,language=english]{acmart}
\AtBeginDocument{%
  }

\copyrightyear{2026}
\acmYear{2026}
\setcopyright{cc}
\setcctype{by}
\acmConference[CF Companion '26 ]{Proceedings of the 23rd ACM International Conference on Computing Frontiers}{May 19--21, 2026}{Catania, Italy}
\acmBooktitle{Proceedings of the 23rd ACM International Conference on Computing Frontiers (CF Companion '26 ), May 19--21, 2026, Catania, Italy}
\acmDOI{10.1145/3801488.3806242}
\acmISBN{979-8-4007-2569-2/2026/05}

\begin{document}

\title{EPAC: The Last Dance}
\subtitle{Invited Paper}

\author{Filippo Mantovani}
\orcid{0000-0003-3559-4825}
\authornote{Corresponding author: <name>.<surname>@bsc.es}
\author{Fabio Banchelli}
\orcid{0000-0001-9809-0857}
\author{Pablo Vizcaino}
\orcid{0000-0002-9253-8275}
\author{Roger Ferrer}
\orcid{0000-0003-3306-8610}
\author{Oscar Palomar}
\orcid{0000-0001-6729-4187}
\author{Francesco Minervini}
\orcid{0000-0001-8558-5690}
\author{Jesus Labarta}
\orcid{0000-0002-7489-4727}
\author{Mauro Olivieri}
\orcid{0000-0002-0214-9904}
  \authornote{Current affiliation: Univ. of Rome Sapienza, Dept. of Information
    Engineering, Electronics, Telecommunications - DIET (Italy)}
\affiliation{%
  \institution{Barcelona Supercomputing Center}
  \country{Spain}
}

\author{Sebastiano Pomata}
\orcid{0000-0003-2377-1273}
\author{Pedro Marcuello}
\orcid{0000-0001-6104-9105}
\author{Jordi Cortina}
\orcid{0000-0002-7201-5880}
\author{Alberto Moreno}
\orcid{0000-0001-8133-0068}
\author{Josep Sans}
\orcid{0000-0003-1983-0680}
\author{Roger Espasa}
\orcid{0009-0008-4885-3631}
\affiliation{%
  \institution{Semidynamics}
  \country{Spain}
}

\author{Vassilis Papaefstathiou}
\orcid{0000-0002-5443-6470}
\author{Nikolaos Dimou}
\orcid{0000-0002-5788-0892}
\author{Georgios Ieronymakis}
\orcid{0000-0001-9776-985X}
\author{Antonis Psathakis}
\orcid{0009-0000-6571-1813}
\author{Michalis Giaourtas}
\orcid{0009-0009-2226-801X}
\author{Iasonas Mastorakis}
\orcid{0009-0001-6279-2002}
\author{Manolis Marazakis}
\orcid{0000-0002-4768-3289}
\affiliation{%
  \institution{Foundation for Research and Technology -- Hellas}
  \country{Greece}
}

\author{Eric Guthmuller}
\orcid{0009-0002-0678-7599}
\author{Andrea Bocco}
\orcid{0000-0002-5862-9819}
\author{Jérôme Fereyre}
\orcid{0009-0004-1519-7333}
\author{César Fuguet}
\orcid{0000-0003-0656-2023}
  \authornote{Current affiliation: Univ. Grenoble Alpes, Inria, CNRS,
    Grenoble INP, TIMA, Grenoble (France)}
\affiliation{%
  \institution{Université Grenoble Alpes, CEA, LIST}
  \country{France}
}

\author{Mate Kovač}
\orcid{0000-0001-7486-4627}
\author{Mario Kovač}
\orcid{0000-0002-8365-7002}
\author{Luka Mrković}
\orcid{0009-0004-4518-5499}
\author{Josip Ramljak}
\orcid{0009-0004-2608-8802}
\affiliation{%
  \institution{University of Zagreb, Faculty of Electrical Engineering and Computing}
  \country{Croatia}
}

\author{Luca Bertaccini}
\orcid{0000-0002-3011-6368}
\author{Tim Fischer}
\orcid{0009-0007-9700-1286}
\author{Frank K. Gurkaynak}
\orcid{0000-0002-8476-554X}
\author{Paul Scheffler}
\orcid{0000-0003-4230-1381}
\author{Luca Benini}
\orcid{0000-0001-8068-3806}
\affiliation{%
  \institution{ETH Z\"urich}
  \country{Switzerland}
}

\author{Bhavishya Goel}
\orcid{0000-0001-9878-4509}
\author{Madhavan Manivannan}
\orcid{0000-0002-9783-8357}
\affiliation{%
  \institution{Chalmers University of Technology}
  \country{Sweden}
}

\author{Tiago Rocha}
\orcid{0009-0005-5609-280X}
\author{Nuno Neves}
\orcid{0000-0003-0628-2259}
\affiliation{%
  \institution{INESC-ID, Instituto Superior Técnico, Universidade de Lisboa}
  \country{Portugal}
}

\author{Jens Krüger}
\orcid{0000-0001-9444-657X}
\affiliation{
  \institution{Fraunhofer ITWM}
  \country{Germany}
}


\begin{abstract}
  This paper presents EPAC, a RISC-V-based accelerator chip developed within the European Processor Initiative (EPI) as part of a multi-year, multi-partner effort to build a European HPC processor ecosystem.
EPAC is implemented in GlobalFoundries 22FDX (GF22FDX) technology, covers an area of 27~mm$^2$ with approximately 0.3 billion transistors, and integrates three distinct RISC-V compute tiles targeting different workload classes:
  VEC, a vector processing tile for double-precision HPC workloads; 
  STX, a many-core tile optimized for stencil and machine learning computations; and 
  VRP, a variable-precision tile for iterative numerical solvers requiring extended floating-point formats. 
All tiles are connected through a Coherent Hub Interface (CHI) based network-on-chip with a distributed L2 cache system and communicate with external memory via a SerDes link.
The chip was taped out in GF22FDX technology and successfully brought up, with all major IP blocks validated.
This paper describes the architecture of each tile and the uncore infrastructure, the integration and physical implementation process, and the board-level bring-up activities.
It also reflects on the engineering and coordination lessons learned from a full chip design effort distributed across academic and industrial partners in Europe.

\end{abstract}




\def\authors{\shortauthors}
\renewcommand{\shortauthors}{Mantovani et al.}

\maketitle

\section{Introduction}



Europe has played an important role in the development of instruction set architectures (ISAs) for embedded systems, especially with Arm. However, it has not reached the same position in high-performance computing (HPC), datacenter, or automotive processors.
In recent years, geopolitical changes and the supply chain issues during the COVID-19 pandemic have shown the risks of relying on non-European technologies. This led to increased efforts to build a European processor ecosystem. The European Processor Initiative (EPI) is one of the main projects in this direction.
EPI is organized in two phases. The first phase (Specific Grant Agreement 1, or SGA1) ran from December 2018 to December 2021 with a budget of €80 million. The second phase (Specific Grant Agreement 2, or SGA2), from January 2022 to March 2026, has a budget of €70 million. The whole initiative was funded 50\% by the EuroHPC Joint Undertaking and 50\% by participating countries (Croatia, France, Germany, Greece, Italy, Netherlands, Portugal, Spain, Sweden, and Switzerland). The total budget assigned to the RISC-V development within EPI (SGA1 and SGA2 together) was 20\% of the total budget (i.e., $\sim$€30 million).

{\bf Dual-Track Architecture Strategy in EPI} --
Modern HPC systems are typically based on a general-purpose CPU combined with one or more accelerators. In practice, CPUs are mainly provided by Intel and AMD, while the accelerator market is dominated by GPU-based solutions, especially from NVIDIA. This setup also defines the common programming model: the CPU runs the operating system and controls execution, while accelerators execute offloaded tasks.
EPI follows this structure, but extends it by combining a general-purpose CPU with multiple accelerators. The CPUs are based on Arm and are developed by SiPearl. On the accelerator side, EPI explores multiple designs, all based on the RISC-V ISA.
In addition, EPI increases flexibility by supporting two modes of operation. Accelerators can be used in a traditional host-device model, but they can also run independently. In this second mode, they are full Linux-capable and can boot and execute workloads without a host CPU. Supporting both models makes the system more flexible than traditional GPU-based approaches.
This paper focuses on the RISC-V developments in EPI, referred to as EPAC (European Processor Accelerator), which is a RISC-V-based accelerator chip including three different approaches to computing acceleration.

{\bf EPAC Vision} --
The EPAC effort is based on three main ideas.
First, all designs use the RISC-V ISA. This supports an open and flexible approach, reduces dependency on proprietary technologies, and avoids vendor lock-in.
Second, EPAC includes multiple accelerator designs (compute tiles) instead of a single solution. This allows exploring different architectural options and sharing knowledge across partners.
Third, EPAC is a full development effort. It includes architecture design, implementation, verification, tape-out, and software development (firmware, compilers, libraries and system software), carried out by a European collaboration of industrial and academic partners (Barcelona Supercomputing Center, Semidynamics, CEA, ETH Zürich, Fraunhofer, FORTH, Chalmers, Extoll, E4 Engineering, IST, Universities of Zagreb, Bologna and Pisa).

%
This paper presents:
{\em i)} the design of three RISC-V compute tiles: VEC, STX, and VRP;
{\em ii)} the development of a system-level uncore infrastructure;
{\em iii)} the integration and tape-out of a full RISC-V accelerator chip (EPAC);
{\em iv)} a coordinated engineering effort across multiple academic and industrial partners.

The remaining part of the document is organized as follows:
Section~\ref{secEpacArch} introduces the EPAC overall system architecture;
Section~\ref{secEngines} details the RISC-V compute tiles;
Section~\ref{secUncore} summarizes the hardware development necessary for a test-chip that are external to the compute tiles (so called ``uncore IPs'');
Section~\ref{secTapeout} presents the tape-out, bring-up and prototyping process;
Section~\ref{secConclusions} shares the final lessons learned and conclusions.

\section{EPAC System Architecture}\label{secEpacArch}

\begin{figure}[!tb]
  \centering
  \includegraphics[width=.97\columnwidth]{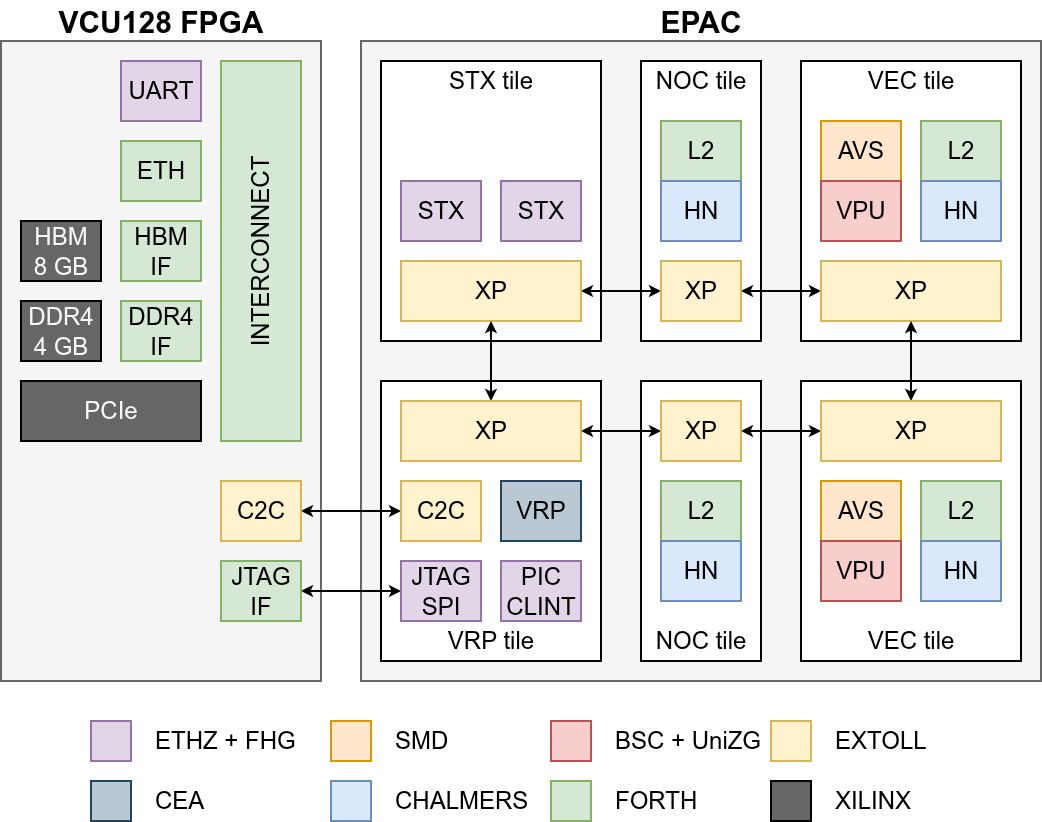}
  \caption{EPAC test-chip block diagram}
  \Description{Block diagram of the EPAC chip connected to a VCU128 FPGA providing DDR4, HBM, PCIe, UART, and Ethernet interfaces. A color-coded legend identifies the partner responsible for each IP block.}
  \label{figEpacBlockDiagram}
\end{figure}

The EPAC chip is implemented in GF22FDX technology. It has an area of 27~mm$^2$ and about 0.3 billion transistors. The chip was submitted for tape-out in October 2022, and bring-up was completed in October 2023. Figure~\ref{figEpacBlockDiagram} shows the high-level block diagram of the system and colors identify the partner developing the technology\footnote{The project adopted Xilinx boards. Xilinx is not part of the project and Xilinx IPs are represented in dark gray in Figure~\ref{figEpacBlockDiagram}.}.

EPAC integrates three types of RISC-V compute tiles: VEC, STX, and VRP. The VEC tile targets traditional HPC workloads, with a focus on double-precision computation. It is based on a general-purpose RISC-V CPU built around the Avispado~\cite{Semidynamics_ProductBrief} in-order core developed by Semidynamics\footnote{A more recent design has been developed including an out-of-order core called Atrevido, part of the portfolio of IPs of Semidynamics.}. The CPU is connected to a vector processing unit (VPU), developed by BSC, through the Open Vector Interface (OVI). The VPU~\cite{minervini2023vitruvius+} supports vector lengths up to 2048 bytes (i.e., 256 double-precision elements) and is compliant with the RISC-V Vector Extension (RVV) version 0.7.1\footnote{This was the ratified version of RVV when the actual development started. A new version of the chip supporting RVV1.0 has been developed and recently taped out.}. The floating-point unit (FPU), developed by the University of Zagreb~\cite{kovavc2023faust}, is integrated inside the VPU.

The STX tile~\cite{FraunhoferITWM_STX} is a RISC-V many-core accelerator designed for machine learning workloads, with a focus on stencil and tensor operations. It includes several small RISC-V cores of two types, Stencil Processing Unit (SPU) and Sparta, developed by Fraunhofer and ETH Zürich. Depending on the workload at runtime, one type of core can be used instead of the other. The tile also includes a 32-bit RISC-V controller that manages the cores and handles memory access through DMA.

The VRP tile~\cite{fuguet2024variable} is a general-purpose RISC-V CPU developed by CEA. It supports variable-precision arithmetic with elements up to 512 bits. This feature is intended for workloads that require higher numerical precision, such as iterative algorithms where convergence can benefit from increased precision.

In addition to the compute tiles, EPAC includes a distributed L2 cache system, with 256~kB of data per slice, developed by FORTH, and a Home Node (HN) developed by Chalmers that manages coherence. Memory accesses from the tiles are routed through the Network-on-Chip (NoC) to slices of the distributed L2 cache and Home Nodes, which identify the data location. If the data is not found in the destination cache slices or other tiles, cache miss requests are then routed via the NoC to external memory.

The NoC, developed by EXTOLL, is based on multiple crosspoints and uses the CHI protocol. EPAC does not include an on-chip memory controller. Instead, it connects to external memory via a 25~GB/s full-duplex chip-to-chip link, also developed by EXTOLL.

Fraunhofer was responsible for the physical design of the chip, while E4 and SECO developed the hardware platforms used to host and test it.

\section{RISC-V Compute Tiles}\label{secEngines}

This section introduces the RISC-V compute tiles developed in the EPI project. These tiles represent three different approaches to accelerating high-performance computing workloads. All three designs have been implemented and taped out in the EPAC test chip. They are not intended to operate together in parallel, but rather to explore different architectural solutions and study their behavior in a real system.

\subsection{VEC tile}
The VEC tile targets traditional HPC workloads, with a focus on double-precision computation. Its goal is to explore a more aggressive design point compared to standard HPC CPUs, while still keeping the programmability and flexibility of a general-purpose processor. In practice, VEC behaves like a RISC-V CPU capable of booting a full operating system, combined with a large vector unit that enables acceleration of compute-intensive workloads. Since it is a general-purpose processor, it can execute any RISC-V binary compliant with RISC-V extensions 
A, B, C, D, F, I, M, V, Zicsr, Zifencei, Zicbom and Zicboz.
A quantitative analysis of this architectural approach is provided in~\cite{vizcaino2023short}.

From an architectural point of view, VEC is built around two main components: a scalar RISC-V core (Avispado, developed by Semidynamics) and a vector processing unit (VPU) developed by BSC. The scalar core is an in-order design capable of booting Linux. One of its key features is the ability to issue a large number of outstanding cache-line requests through the Gazillion unit, which acts as a Miss Status Holding Register (MSHR) based mechanism for handling many concurrent memory requests. The core is connected to the VPU through the Open Vector Interface (OVI), an open hardware interface defined by Semidynamics~\cite{Semidynamics_OVI}.

\begin{figure}[!tb]
  \centering
  \includegraphics[width=.97\columnwidth]{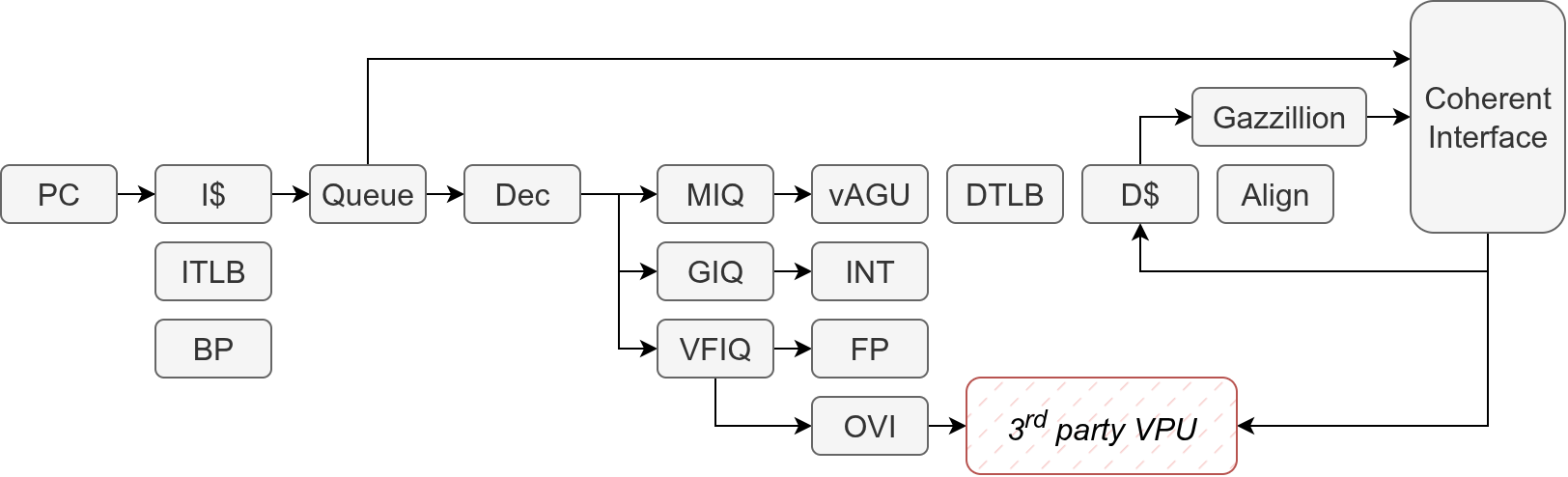}
  \caption{Avispado + VPU block diagram}
  \Description{Pipeline diagram of the Avispado scalar core. Instructions flow from the program counter through the instruction cache and decode stage into three dispatch queues: MIQ for memory operations, GIQ for integer operations, and VFIQ for vector and floating-point operations. The VFIQ connects through the OVI interface to a third-party VPU.}
  \label{figAvispadoBlockDiagram}
\end{figure}

The internal structure of the core follows a typical general-purpose processor design as depicted in Figure~\ref{figAvispadoBlockDiagram}. Instructions are decoded and dispatched into different queues depending on their type: memory operations, integer operations, and vector/floating-point operations. The memory queue includes the Gazillion unit, while the vector/floating-point queue is connected to the VPU through the OVI. The core includes a 16~kB instruction cache and a 32~kB data cache, supports unaligned memory accesses in hardware, and supports vector memory operations through a dedicated store queue. The scalar core directly manages coherence through the CHI protocol, making the tile compatible with the rest of the EPAC system without additional integration complexity.

The VPU can be seen as a pipelined SIMD unit operating on vector registers up to 2048 bytes, corresponding to 256 double-precision elements. It includes 40 physical vector registers, compared to the 32 registers defined by the RISC-V standard, allowing a limited form of register renaming with up to 8 additional registers before spilling is required. The VPU has eight parallel functional units, each including FAUST\footnote{Part of Proactive Compute, \url{https://proactivecompute.com}}, a pipelined vector floating-point unit developed by the University of Zagreb~\cite{kovavc2023faust}. Because of this structure, a vector arithmetic instruction operating on a full vector requires 32 cycles, plus a small decode overhead (around 3 cycles), as the eight functional units process eight elements per cycle, completing all 256 elements in 32 cycles. Memory operations may have higher and variable latency.

The VPU is vector-length agnostic, following the RISC-V Vector Extension (RVV 0.7.1). Software can set the active vector length up to the hardware maximum, allowing loops of arbitrary size to be processed without scalar tail handling. More details on the VPU design can be found in~\cite{minervini2023vitruvius+}.

From a programming point of view, VEC follows a standard HPC workflow. Within EPI, an LLVM-based toolchain has been developed to support automatic vectorization for C, C++, and Fortran code~\cite{LLVM_EPI}. This is the main approach to maintain code portability. When more control is needed, developers can use guided vectorization through pragmas following the OpenMP SIMD syntax. For low-level optimization, intrinsic functions (built-ins) are also available, allowing direct mapping to vector instructions at the cost of portability. As a last option, inline assembly can be used.

Execution of code on VEC follows a typical HPC environment: applications are compiled with the LLVM toolchain, linked with either vectorized or generic RISC-V libraries, and executed through job scheduling systems such as Slurm. The development flow also includes software and hardware emulation tools that allow analysis of vectorization efficiency, for example to detect underutilization of the available vector length or missed vectorization opportunities by the compiler. This set of tools and platforms is referred to as the Software Development Vehicle (SDV) environment~\cite{mantovani2023software,vizcaino2025designing,banchelli2025risc} and has been used for quantifying the benefit of having a large vector architecture as well as for analyzing scientific codes running on EPAC. More details can be found in~\cite{banchelli2026exploring,banchelli2024batched,torres2024co,vizcaino2024graph,blancafort2024exploiting}.
\subsection{STX tile}

Stencil computations and many tensor operations share common characteristics: they rely on regular multi-dimensional data access patterns, have high memory bandwidth requirements with limited data reuse, and involve a large number of simple floating-point operations per grid point. These workloads are typically bandwidth-bound and do not benefit from deep cache hierarchies or complex out-of-order cores. In addition, general-purpose processors introduce overhead due to address generation, cache management, and control logic, even when access patterns are known in advance.

The STX tile is designed to address these limitations. It exploits regular access patterns using hardware address generation units and hardware loop controllers, replaces cache-based memory hierarchies with scratchpad memories managed through DMA, and integrates domain-specific compute units designed for stencil and tensor workloads.

STX is a domain-specific accelerator targeting stencil computations, machine learning, and deep learning workloads. Typical use cases include structured grids in 2D and 3D, fixed stencil patterns such as 7-point or 27-point stencils, iterative time-stepping algorithms (e.g., diffusion or wave propagation), and tensor operations such as convolutions, pooling, and matrix-matrix multiplications. The design also supports mixed precision and experimental number formats such as POSIT.

Within EPI, STX originates from the integration of two initially separate accelerator concepts: a stencil accelerator and a machine learning accelerator. These were merged into a single design by Fraunhofer, ETH Zürich, and the University of Bologna, resulting in the STX (stencil/tensor accelerator), with an optional extension called the Stencil Processing Unit (SPU) to further improve performance on stencil workloads.

From an architectural point of view, STX is built around a lightweight RISC-V core called Snitch~\cite{zaruba2020snitch}. This is a small 32-bit core coupled with a 64-bit floating-point unit with SIMD support. Unlike traditional designs, where the integer pipeline is heavily used for address generation, STX offloads this functionality to Stream Semantic Registers (SSRs). SSRs allow memory streams to be mapped directly to floating-point registers, bypassing the load/store unit and reducing contention. This enables very efficient data feeding to the FPU. For example, in a simple AXPY kernel, the FPU can be sustained at one operation per cycle using only a small number of registers.

The architecture also includes Floating-Point Repetition (FREP), which allows the hardware to repeat floating-point instruction sequences without re-fetching them from the instruction cache. This reduces both instruction overhead and energy consumption in tight loops.

A typical STX cluster includes multiple Snitch compute cores (typically eight), along with a dedicated Snitch core enhanced with DMA capabilities to manage data movement. The system can be extended with SPU units, which are optional co-processors developed by Fraunhofer and optimized for stencil workloads with static access patterns and local data dependencies. These units are designed through a hardware-software co-design approach and provide additional performance for stencil kernels.

\begin{figure}[!tb]
  \centering
  \includegraphics[width=.97\columnwidth]{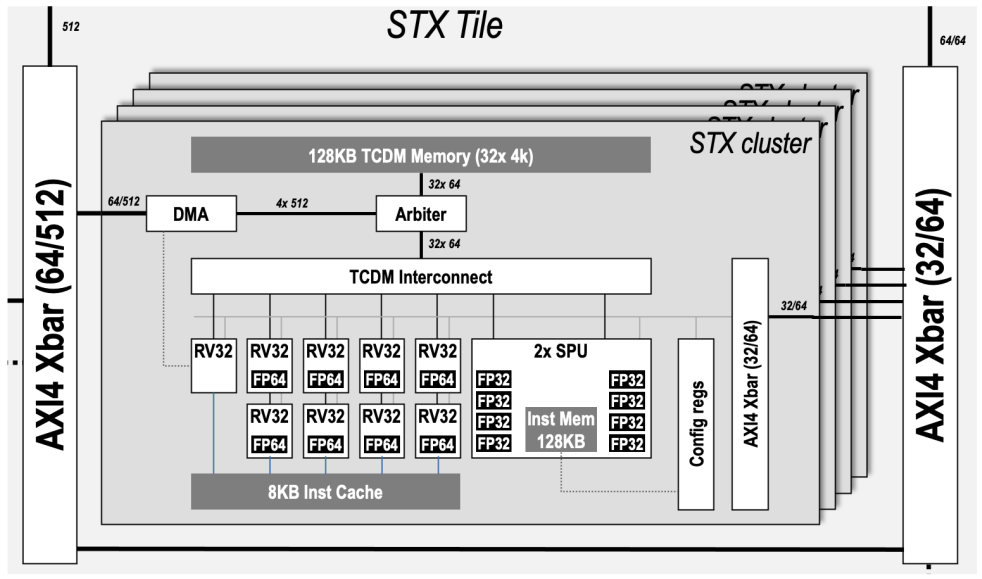}
  \caption{STX block diagram}
  \Description{Block diagram of an STX tile showing four stacked STX clusters connected through AXI4 crossbars. Each cluster contains 128 kB of TCDM (Tightly-Coupled Data Memory) scratchpad memory, a DMA engine with arbiter, a TCDM interconnect, eight RV32 Snitch compute cores each with an FP64 unit, two SPU co-processors with FP32 units and 128 kB instruction memory, an 8 kB instruction cache, and configuration registers. External interfaces use AXI4 crossbars with 64/512-bit and 32/64-bit data widths.}
  \label{figStxBlockDiagram}
\end{figure}

STX follows a modular design. Each accelerator consists of multiple clusters (typically four), and each cluster includes 4 to 16 Snitch compute cores, one DMA controller core, and optionally up to four SPU units. All units share a local scratchpad memory (64-256 kB), which is explicitly managed through DMA. Figure~\ref{figStxBlockDiagram} shows a block diagram of an STX tile implemented in EPAC. In a representative configuration with four clusters and eight compute cores per cluster running at 1~GHz, the system can reach up to 64~GFLOPS in double precision per tile. In practice, high utilization of the floating-point units has been observed for machine learning workloads.

More generally, by relying on a simpler architecture without large cache hierarchies or complex out-of-order logic, STX is a strong candidate for achieving higher energy efficiency compared to general-purpose or vector-based designs.

From a programming point of view, STX supports OpenMP-based offloading. Workloads can be offloaded from an Arm-based general-purpose processor or from a 64-bit RISC-V core within EPAC. Both GCC- and LLVM-based toolchains are supported.
  

\subsection{VRP}

The VRP tile~\cite{fuguet2024variable,guthmuller2025variable} is designed to efficiently execute floating-point computations at higher precision than standard double precision, while allowing the precision to be adapted at runtime based on the numerical requirements of the application. The main target use case is iterative linear solvers, such as Krylov methods (e.g., CG, BiCG, PCG), where increasing precision can reduce rounding errors~\cite{durand2022accelerating,hoffmann2024stabilizing}, improve convergence, or enable convergence for ill-conditioned systems.

Unlike software-based multi-precision libraries such as MPFR, which rely on sequences of operations on general-purpose CPUs, VRP provides dedicated hardware support for variable-precision floating-point computation. This includes a custom RISC-V ISA extension (Xvpfloat~\cite{guthmuller2024xvpfloat}), a dedicated floating-point unit, and support for extended IEEE 754 formats in memory.

The VRP tile is built around a modified 64-bit RISC-V CVA6 core, a 16~kB instruction cache, a 32~kB high-performance data cache (HPDcache), and a variable-precision floating-point unit (VPFPU). The tile is connected to the EPAC system through a CHI coherent interface, allowing it to access shared memory and operate within the same address space as the other tiles and the host. This enables a standard host-device execution model, where data is prepared by the host and processed by VRP with cache coherence maintained.

The CVA6 core is a single-issue, in-order design capable of running a full operating system such as Linux. It is extended to support the Xvpfloat ISA, a dedicated floating-point register file, and additional control logic to manage operations with variable latency. The Xvpfloat extension introduces custom instructions for arithmetic, comparison, and data movement on variable-precision floating-point values, along with conversion between standard formats (float/double) and extended formats.

The ISA defines 32 logical floating-point registers (P-registers), which can store values up to the maximum supported precision, and a set of environment registers that configure the precision at runtime. These registers control the number of bits used for the significand and exponent, the rounding mode, and the memory format. This design separates the representation in memory from the internal computation format, allowing the precision to be adjusted without recompiling the code.

At the hardware level, the implementation includes a larger physical register file with 64 entries and a rename mechanism that maps logical P-registers to physical registers. Together with a scoreboard, this allows limited overlap of independent operations, even in a single-issue core, which helps hide the increased latency of high-precision arithmetic and memory operations.

The VPFPU supports floating-point formats with a significand up to 512 bits and an exponent up to 18 bits. To avoid the cost of a full-width 512-bit datapath, it uses a chunk-based architecture. Some operations, such as normalization and leading-zero detection, are performed at full width, while arithmetic operations are implemented using smaller units (e.g., 128-bit adders and 64-bit multipliers) that iterate over multiple chunks depending on the selected precision.

The VPFPU is organized into separate pipelines for addition, subtraction, multiplication, move/compare/convert, and load/store operations. Latency and throughput scale with the selected precision, but multiple pipelines allow independent instructions to be executed in parallel. This helps compensate for the higher latency of extended-precision operations and enables sustained throughput close to one instruction per cycle for suitable instruction mixes.

The memory system is designed to match the bandwidth requirements of these workloads. The HPDcache supports access patterns typical of BLAS-like kernels, with strided accesses and multiple outstanding misses (up to 128). It also includes a programmable prefetcher and can deliver up to 16 bytes per cycle to the VPFPU. As a result, the system is typically limited by memory bandwidth rather than compute capability, which is consistent with the target workloads.

In memory, VRP uses extendable IEEE 754 formats, where the total number of bits, as well as exponent and significand sizes, can be configured at runtime. This allows selecting different trade-offs between memory footprint and numerical accuracy, for example using 128, 256, or 512-bit representations depending on the application.

The typical usage model follows a host-device approach. The host loads the program, runtime, and data into memory, configures the VRP through control registers, and starts execution. The VRP runs a RISC-V binary using specialized libraries (e.g., VBLAS) to operate on extended-precision data types.

At the end of execution, the VRP signals completion to the host through either memory or inter-processor interrupts. The precision can be configured at runtime through environment registers, allowing adaptive strategies that balance performance, energy consumption, and numerical stability without recompilation.

\section{Uncore Architecture}\label{secUncore}

The ``Uncore'' of EPAC consists of a NoC that uses the AMBA~5 CHI protocol to interconnect the compute tiles with a distributed shared L2 cache which is coupled with a coherence Home Node (HN). It also includes a Chip-to-Chip (C2C) link that extends the CHI network off-chip to enable connectivity with memories and external I/O devices.

The shared L2 cache and the tightly-coupled coherence home comprise multiple slices that operate with the AMBA~5 CHI protocol. The design is modular and instantiated on several nodes of the CHI NoC to create multiple distributed slices, each responsible for a subset of the coherent system address space. 
Each L2 cache slice is 256~kB, eight-way set-associative, non-blocking, write-back cache with pseudo-LRU replacement policy, with support for 128 outstanding transactions, 64 misses and 64 evictions. The design is fully-pipelined and optimized for throughput in order to serve the bursty patterns of long-vector memory accesses. 
The L2 cache data arrays in each slice have an internal data-path width of 512-bit (64-byte) in order to support up to one cache-line per cycle per slice. 
It also supports cache management operations, non-temporal hints and direct memory transfers (DMT) for certain types of requests to allow bypassing the cache, and allows programmable address interleaving modes. 
The design also includes an Atomic ALU that allows ``far-atomic'' operations to be performed directly inside the L2 cache. Further developments include support for error correction and detection (SECDED) and error reporting interfaces.

The HN is a fully mapped directory cache coherence controller responsible for tracking sharers at the cache-line granularity and for ensuring coherence among the copies in the private L1 caches. The HN supports MESI-like cache coherence protocol and is compatible with the AMBA~5 CHI specification. The HN implementation is fully pipelined with separate request and response paths. This allows HN to service requests and responses in parallel with the capability of handling a single request and response every cycle. The HN uses tags maintained by the inclusive L2 cache bank and stores a fully-mapped directory of presence bits and state associated with each tag entry to track and manage the cache-line sharers. 

The high-performance CHI NoC transfers data from main memory into the L2 caches, from the L2 caches into the cores and between the cores. The basic building block of the NoC is the crosspoint (XP) which includes four links to connect to other XPs and two ports for devices like the different compute elements or the L2/HN. The NoC has dedicated channels with credit-based flow control for requests, responses, data, and snoops with multicast support. 
The NoC is configured as a 2D Mesh with dimension order routing. Each endpoint can send one flit in each channel every clock cycle. A complete cache-line of 512 bits can be transferred to and from each port of the NoC every clock cycle. Running at a frequency of 1~GHz, the NoC offers a bandwidth of 64~GB/s per port per direction on the data channels.

The EPAC chip does not include memory controllers and uses the C2C link to extend the CHI NoC off-chip and enable access to external DRAM memory and I/O devices (via an FPGA as described in Section~\ref{secTapeout}). The C2C link uses 8 SerDes lanes, each capable of operating up to 25~Gb/s. Assuming zero link and packet overhead, when the SerDes lanes operate at the maximum rate, the C2C link can provide a maximum bandwidth of 200~Gb/s per direction (25~GB/s), thus an aggregate bandwidth of 400~Gb/s (50~GB/s) 
which is adequate to saturate a DDR4 DRAM memory channel. The CHI NoC flits are encapsulated in packets with CRC checksums, and the C2C supports link-level retransmission to ensure reliable communication in the presence of bit errors.

\section{Implementation and Tapeout}\label{secTapeout}

The EPAC test chip was implemented in GlobalFoundries 22FDX Fully Depleted Silicon-On-Insulator (FDSOI) technology, using a 10-layer metal stack. IP selection involved benchmarking two standard cell and memory library vendors (Invecas and Synopsys); the Invecas GF22FDX 8-track library was selected for its better power-performance trade-off. The 8-track variant was preferred over the 12-track option to limit power dissipation. The remaining IP components included Invecas 1.8~V GPIO cells, high-density SRAM compilers, and custom SerDes and LVDS IPs from EXTOLL. The chip was packaged in a flip-chip configuration with 150~$\mu$m bump pitch. 

\begin{figure}[!tb]
  \centering
  \includegraphics[height=3.4cm,keepaspectratio]{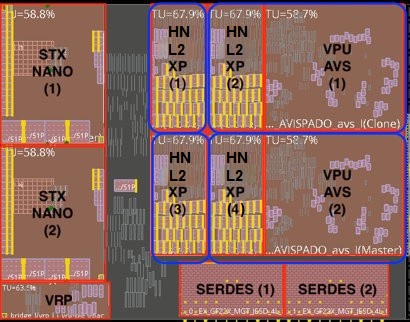}\hfill%
  \includegraphics[height=3.4cm,keepaspectratio]{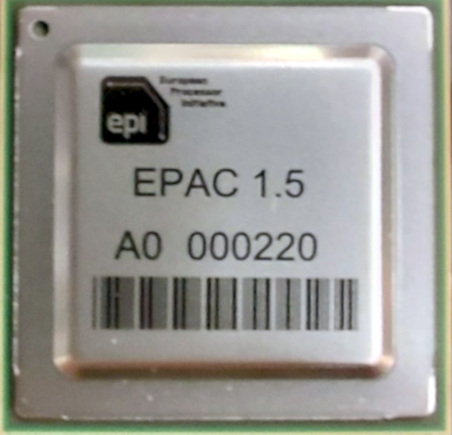}
  \caption{Left: Floorplan with area of EPAC components. Right: Photo of the EPAC test-chip.}
  \Description{Two side-by-side images. Left: color-coded floorplan of the EPAC chip showing the placement of STX partitions, VRP partition, Home Node and L2 cache blocks with crosspoints, Avispado/VPU partitions and SerDes macro blocks. Right: photograph of the fabricated EPAC die.}
  \label{figEpacPic}
\end{figure}

Integrating RTL from multiple partner organizations introduced several practical challenges (e.g., module renaming, tcl scripting, etc.). 
Code quality was assessed using the Cadence JasperGold linter.
Parameterized SystemVerilog interfaces, while convenient for front-end design, were found to be incompatible with the hierarchical backend flow
so they were replaced with fixed interfaces at all block boundaries.
Code freeze was reached in January 2022.
Given the scale of the design (over 14M logic cells and 991 memory macros) a hierarchical implementation flow was adopted using the Cadence tool suite: Genus for synthesis, Innovus for place and route, Tempus for static timing analysis, Voltus for power rail analysis, Quantus for parasitic extraction, and Conformal for logical equivalence checking (LEC). Physical verification was performed with Siemens Mentor Calibre against GF22FDX foundry rules.

The design was split into three implementation partitions: the STX tile (1.86M cells), the VEC tile (2.26M cells), and the VRP tile (532K cells). Synthesis was overconstrained to 1.25~GHz to account for on-chip variation, against a 1~GHz operational target. Retiming was applied to meet timing on critical paths.
Design-for-testability (DFT) coverage included full-scan, scan compression, and Programmable Memory Built-In Self-Test (PMBIST) for all memory macros, accessed via a standard JTAG interface.
%
Iterative floorplan optimization reduced the chip area from an initial estimate of over 30~mm$^2$ to a final implementation area of 26.97~mm$^2$ (27.29~mm$^2$ including scribe line). All three partitions achieved clean Design Rule Check (DRC) and Layout vs. Schematic (LVS) at signoff. Static IR-drop at the assembled top level reached 47~mV under worst-case conditions, within acceptable limits.
Figure~\ref{figEpacPic} (left) shows the floorplan with components identified.
The final Graphic Database System (GDS) was submitted to GlobalFoundries on 10 October 2022, following a pre-tape-out verification run during August.
The chip operates at 768~MHz under worst-case conditions (SS corner, 0.72~V, 125~°C) and up to 1234~MHz at typical conditions (TT, 0.80~V, 85~°C). Silicon samples arrived on 17 February 2023.
Figure~\ref{figEpacPic} (right) shows a picture of the chip.

The chip was packaged in an FCBGA format (22$\times$22 balls, 1~mm pitch) and 
mounted on a custom daughterboard designed by E4 and SECO, which interfaces with a Xilinx VCU128 FPGA board.
Bring-up was conducted in parallel by two teams: FORTH handled core IP validation (NoC, VPU, L2HN, interrupt controllers), while EXTOLL focused on the SerDes link.
A custom bring-up shell provided interactive access to on-chip cores via an emulated JTAG console, enabling early testing before full SerDes connectivity was available.
The validation sequence covered JTAG/SPI register access, SRAM read/write patterns, inter-tile CHI and AXI connectivity, cache coherency and snoop delivery, VPU vectorized benchmarks (DGEMM and Stream), and CLINT interrupt delivery across micro-tiles, all passing successfully.
In the prototype where EPAC is connected with the FPGA (see Figure~\ref{figEpacBlockDiagram} for details), the C2C link was safely demonstrated with aggregate bandwidth of 20~GB/s. This allowed interfacing with DDR4 and HBM on the FPGA and enabled workloads with multi-GByte memory footprints. After the full bring-up, EPAC successfully booted Ubuntu 22.04 LTS using Linux kernel v5.7 with vector support. The EPAC chip demonstrated stability, handling standard GUIs and executing long HPC benchmarks like LINPACK or other scientific applications without issues.

\section{Conclusion}\label{secConclusions}

EPAC demonstrates that a full chip design effort, from architecture to tape-out and bring-up, can be successfully carried out by a distributed European consortium combining academic and industrial partners. 
The chip integrates three distinct RISC-V compute tiles (VEC, STX, and VRP) in a single silicon prototype, each targeting a different class of HPC workloads: vector-parallel double-precision computation, stencil and machine learning acceleration, and extended-precision iterative solvers. 
Rather than optimizing a single design point, EPAC deliberately explores architectural diversity, providing a concrete basis for comparing different acceleration strategies under the same system-level constraints and within the same memory and interconnect infrastructure.

The engineering process revealed a set of challenges that are common to large-scale collaborative chip design but are rarely documented in detail.
Integrating RTL from multiple organizations required non-trivial consolidation work at the system level, including interface standardization, naming conflict resolution, and toolchain alignment between front-end and back-end flows.
The hierarchical physical implementation, covering over 14 million logic cells across three independently implemented partitions, demanded tight coordination between physical design, timing closure, and DFT insertion, activities that had to proceed in parallel with ongoing RTL changes from partner teams.
These challenges are not unique to EPAC, but the scale and distributed nature of the effort made them particularly visible.

From an architectural standpoint, the three tiles confirm that there is no single best approach for HPC acceleration.
VEC provides the most general solution, capable of running a full software stack and benefiting from compiler-driven vectorization, at the cost of higher design complexity.
STX trades generality for efficiency: its simpler datapath, scratchpad-based memory hierarchy, and hardware-level data streaming make it a strong candidate for energy-efficient execution of regular workloads.
VRP addresses a more specialized niche of scientific computing (extended-precision arithmetic for ill-conditioned numerical problems) and shows that dedicated hardware support for variable precision is both feasible and practical in a RISC-V framework.

Looking ahead, the EPAC effort provides both a validated silicon platform, a portfolio of silicon-proven IPs and a body of engineering experience that directly feeds into future European processor developments.
The software infrastructure (compilers, libraries, emulation tools) developed alongside the hardware represents an asset that extends beyond the chip itself.
The main open challenge remains the scalability of the multi-partner design model: as chip complexity grows, the coordination overhead between organizations becomes a first-class design constraint that requires explicit planning, not just technical solutions.

\section{Acknowledgments}
This project has received funding from the EuroHPC-JU under European Processor Initiative FPA N.~800928, SGA N.~826647 (EPI-SGA1) and SGA N.~101036168 (EPI-SGA2). The JU receives support from the European Union's Horizon 2020 research and innovation programme and from Croatia, France, Germany, Greece, Italy, Netherlands, Portugal, Spain, Sweden, and Switzerland.

\bibliographystyle{unsrtnat}
\bibliography{sample-base}

\end{document}